\documentclass[12pt]{iopart}

\usepackage{iopams}
\usepackage{multicol,mathrsfs,bm}
\usepackage{graphicx, graphics}
\usepackage{hyperref}
\usepackage{slashed}
\usepackage{fancyhdr}

\def\k{{\bf k}}

\renewcommand{\not}{\slashed}
\renewcommand{\overline}{\bar}

\newcommand{\gsim}{\geq}
\def\gappeq{\mathrel{\rlap {\raise.5ex\hbox{$>$}} {\lower.5ex\hbox{$\sim$}}}}
\def\lappeq{\mathrel{\rlap{\raise.5ex\hbox{$<$}} {\lower.5ex\hbox{$\sim$}}}}
\def\be{\begin{equation}} \def\ee{\end{equation}}
\def\bea{\begin{eqnarray}} \def\eea{\end{eqnarray}}
\def\bq{\begin{quote}} \def\eq{\end{quote}}
\def\bc{\begin{center}} \def\ec{\end{center}}

\begin{document}

 \begin{flushright}
\textbf{JCAP} 08 (2007) 002\\
\small {DFPD-07/TH/04}\\
{\small MIT-CTP-3817}
 \end{flushright} 
 
\title{ Quantum Boltzmann Equations  and Leptogenesis}

\author{Andrea De Simone $^1$ and Antonio Riotto $^{2, 3}$} 

\address{$^1$ Center for Theoretical Physics,\\
Massachusetts Institute of Technology, Cambridge, MA 02139, USA}

\address{$^2$ INFN, Sezione di Padova, Via Marzolo, 
8 - I-35131 Padua, Italy}
        
\address{$^3$ D\'epartement de Physique Th\'eorique,\\ Universit\'e de
Gen\`eve, 24 Quai Ansermet, Gen\`eve, Switzerland}
\eads{\mailto{andreads@mit.edu} and \mailto{antonio.riotto@pd.infn.it}}

\begin{abstract}
The closed time-path  formalism is a 
powerful Green's function formulation to describe 
non-equilibrium phenomena in field theory and 
it leads to a complete non-equilibrium quantum 
kinetic theory. We make use of this  formalism to write down the
set of quantum Boltzmann equations relevant for leptogenesis.  
They manifest memory effects and off-shell corrections. In particular,
memory effects lead to a time-dependent CP asymmetry whose   
value at a given instant of time depends upon the 
previous history of the system. This result is particularly
relevant when the asymmetry is generated by the decays of nearly 
mass-degenerate heavy states, as in resonant or soft leptogenesis.
\end{abstract}

\maketitle


\pagestyle{fancy}
\def\thefootnote{\arabic{footnote}}
\setcounter{footnote}{0}
\setcounter{page}{1}

\fancyhead{}
\fancyfoot[C]{\thepage}


\section{Introduction}

Thermal leptogenesis~\cite{fy,lept,ogen,work} is a well-motivated scenario
for the production of the baryon asymmetry in the early Universe.
It takes place through the
decay of heavy right-handed (RH)  Majorana neutrinos. 
The out-of-equilibrium decays occur violating
lepton number and CP, thus satisfying Sakharov's
conditions~\cite{sakharov}. In grand unified theories (GUT) the
masses of the heavy Majorana neutrinos masses are typically smaller
than the scale of unification of the electroweak and strong
interactions, $M_{\rm GUT} \sim 10^{16}$~GeV, by a few to
several orders of magnitude. This range coincides with the range of
values of the heavy Majorana neutrino masses required for a
successful thermal leptogenesis. 
Being RH neutrinos   a key ingredient in the formulation of
the well-known seesaw mechanism~\cite{seesaw},  
which explains why neutrinos are
massive and mix among each other and  why they turn out to be
much lighter than the other known fermions
of the Standard Model (SM), thermal leptogenesis has been the
subject of intense research activity in the last few years.
For instance,  flavour effects 
have been recently investigated in  detail (and shown to be relevant)
\cite{Barbieri99,endoh,davidsonetal,nardietal,dibari,
davidsonetal2,antusch,silvia1,Branco:2006ce,aat,silvia2,adsar,vives,
Engelhard:2006yg}
including the quantum oscillations/correlations of the asymmetries
in lepton flavour space~\cite{davidsonetal,adsar}.
The interactions related to the charged Yukawa couplings enter in the
dynamics by inducing nonvanishing 
 quantum oscillations among the lepton asymmetries
in flavour space \cite{davidsonetal}. 
Therefore the lepton asymmetries must be
represented as a matrix  in flavour space, 
the diagonal elements are the  flavour asymmetries, and the off-diagonals
encode the quantum correlations. The off-diagonals 
should decay away when the charged Yukawa couplings start 
mediating very fast 
processes. For instance, at temperatures $T\sim 10^{12}$ GeV, the interactions
mediated by the tau-Yukawa coupling  enter in equilibrium and the
tau-flavour becomes distinguishable from the muon and the electron flavour.
A  full treatment of this transition  
based on the quantum, rather than classical, 
 Boltzmann equations is suitable to properly describe 
all the physical effects. 

Mainly, but not only,  motivated by the impact of flavour effects,
in this paper we set the stage for the study of the dynamics 
of thermal leptogenesis by means of 
quantum Boltzmann equations (for a previous study, see \cite{buchmuller}). 
They are obtained 
starting from  the  non-equilibrium quantum field theory based on the 
Closed Time-Path (CTP) formulation. We will see that the resulting 
kinetic equations describing the evolution of the
lepton asymmetry and the RH neutrinos are non-Markovian and 
present memory effects. In other
words, differently from the classical approach where every scattering
in the plasma is independent from the previous one, the particle abundances
at a given time depend upon the history of the system. The more
familiar energy-conserving delta functions are replaced by retarded
time integrals of time-dependent kernels and 
 cosine functions whose arguments are the energy involved
in the various processes. Therefore, the non-Markovian kinetic equations
include the contribution of coherent processes throughout the history
of the kernels. 

We will see that 
one immediate consequence of the CTP approach to thermal
leptogenesis is that the CP asymmetry is a function of time
and its value at a given
instant depends upon the previous history of the system. 
Furthermore, in the quantum approach,  the 
 relaxation times are expected to be typically 
longer than the one dictated by the classical approach. 
Particle number densities are replaced by 
Green's functions which  are subject both to exponential
decays  and to an oscillatory behaviour. This  restricts the range of time 
integration for the scattering terms and leads to a decrease of the wash-out
rates. This is a well-established fact in nuclear collisions \cite{dan}.
If the time range of the kernels are shorter than the
relaxation time of the particles abundances, the solutions to the
quantum and the classical
Boltzmann equations differ only by terms of the order of the ratio
of the timescale of the kernel to the relaxation timescale of the
distribution. In thermal leptogenesis this is typically the case. However, 
there are situations where this does not happen. For instance, in the
case of resonant leptogenesis and soft leptogenesis, 
two RH (s)neutrinos $N_1$ and $N_2$ 
 are almost degenerate
in mass and the CP asymmetry from the decay of the first RH $N_1$ 
is resonantly enhanced if the mass difference
$\Delta M=(M_2-M_1)$
is of the order of the decay rate of the second RH neutrino
$\Gamma_{N_2}$ . The typical timescale
to build up coherently the CP asymmetry is of the order of $1/\Delta M$, which
can be larger than the timescale $\sim 1/\Gamma_{N_1}$ 
for the change of the abundance of the
$N_1$'s. This tells us that the reduction of the quantum Boltzmann
equations to the classical ones should  not be taken for granted. 

The paper is organized as follows. In Section \ref{ctpform} we provide the basics 
of the CTP
 formulation of non-equilibrium quantum field theory, 
we set the notations and state the main results which will be used throughout the paper. 
The quantum Boltzmann equation describing the time evolution of the lepton 
number asymmetry is formally derived in Section \ref{qbegeneral}, in terms of propagators and 
self-energies of the leptons. Sections \ref{qberhneutrino} and \ref{qbeleptonasym} 
are devoted to write down more explicitly 
the quantum Boltzmann equations for the abundance of right-handed neutrinos 
and the lepton asymmetry, respectively. The familiar classical equations are then 
easily recovered  as a limit case. Section \ref{qbeleptonasym} also contains an analysis of the time-dependent CP asymmetry  obtained in this more general setup. Concluding remarks are in Section \ref{concl}.


\section{The CTP formalism for non-e\-qui\-lib\-ri\-um quantum field theory}
\label{ctpform}

\begin{figure}[t]
\centering
\includegraphics{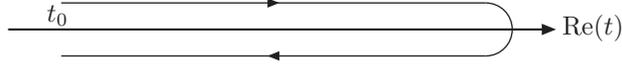}
\caption{The complex time contour for the CTP formalism.}
\label{contour}
\end{figure}

We first  briefly present  some of the  basic 
features of the  non-equilibrium quantum field theory based on the 
CTP formalism, also known as Schwinger-Keldysh formalism \cite{sk}.
The interested reader is referred to the excellent review by Chou   
{\it et al.} \cite{chou} for a more exhaustive discussion. 
Since  we
need the time evolution of the particle asymmetries with definite 
initial conditions and not
simply the transition amplitude of particle reactions, 
the ordinary equilibrium quantum field theory at finite temperature   
is not the appropriate tool. 
The most appropriate extension of the field theory
to deal with non-equilibrium phenomena amounts to generalizing
 the time contour of
integration to a closed time-path. More precisely, the time integration
contour is deformed to run from $t_0$ to $+\infty$ and back to
$t_0$ (see Fig.~\ref{contour}). 
The CTP  formalism  is a powerful 
Green's function
formulation for describing non-equilibrium phenomena in field theory.  It
allows to describe phase-transition phenomena and to obtain a
self-consistent set of quantum Boltzmann equations.
The formalism yields various quantum averages of
operators evaluated in the in-state without specifying the out-state. 
On the contrary, the ordinary quantum field theory 
yields quantum averages of the operators evaluated  
with an in-state at one end and an out-state at the other. 

For example, because of the time-contour deformation, the partition function 
in the in-in formalism for a complex scalar field is defined to be
\begin{eqnarray}
Z\left[ J, J^{\dagger}\right] &=& {\rm Tr}\:
\left[ \mathcal{T}\left( {\rm exp}\left[i\:\int_C\:\left(J(x)\phi(x)+
J^{\dagger}(x)\phi^{\dagger}(x) \right) d^4 x\right]\right)\rho\right]\nonumber\\
&=& {\rm Tr}\:\left[ \mathcal{T}_{+}\left( {\rm exp}\left[ i\:\int\:
\left(J_{+}(x)\phi_{+}(x)+J^{\dagger}_{+}(x)\phi^{\dagger}_{+}(x) \right) d^4 x\right]\right)
\right.
\nonumber\\
&\times&\left.  \mathcal{T}_{-}\left( {\rm exp}\left[
      -i\:\int\:\left(J_{-}(x)\phi_{-}(x)+J^{\dagger}_{-}(x)\phi^{\dagger}_{-}(x)
      \right) d^4 x\right]\right) \rho\right],
\end{eqnarray}
where $C$ in the integral denotes that the time 
integration contour runs from $t_0$ to plus infinity 
and then back to $t_0$ again. The symbol $\rho$ 
represents the initial density matrix and the fields are in 
the Heisenberg picture  and  defined on this closed time-contour (plus and minus subscripts refer to the positive and negative directional branches of the time path, respectively). The time-ordering operator along the path is the standard one ($\mathcal T_+$) on the positive branch,  and the anti-time-ordering ($\mathcal T_-$) on the negative branch.
As with the Euclidean-time formulation, scalar (fermionic) fields $\phi$ are
still periodic (anti-periodic) in time, but with
$\phi(t,\vec{x})=\phi(t-i\beta,\vec{x})$, $\beta=1/T$.
The temperature $T$ appears   due to boundary
condition, and time is now  explicitly present in the integration
contour.

We must now identify field
variables with arguments on the positive or negative directional
branches of the time path. This doubling of field variables leads to
six  different real-time propagators on the contour \cite{chou}.  These six
propagators are not independent, but using all of them simplifies the 
notation. 
For a generic charged  scalar field $\phi$ they are defined as 
\begin{eqnarray}
\label{def1}
G_{\phi}^{>}\left(x, y\right)&=&
-G_{\phi}^{-+}\left(x, y\right)=
-i\langle
\phi(x)\phi^\dagger (y)\rangle,\nonumber\\
G_{\phi}^{<}\left(x,y\right)&=&-G_{\phi}^{+-}\left(x, y\right)=-i\langle
\phi^\dagger (y)\phi(x)\rangle,\nonumber\\
G^t _{\phi}(x,y)&=& G_{\phi}^{++}\left(x, y\right)=
\theta(x,y) G_{\phi}^{>}(x,y)+\theta(y,x) 
G_{\phi}^{<}(x,y),\nonumber\\
G^{\overline{t}}_{\phi} (x,y)&=& G_{\phi}^{--}\left(x, y\right)=
\theta(y,x) G_{\phi}^{>}(x,y)+
\theta(x,y) G_{\phi}^{<}(x,y), \nonumber\\
G_{\phi}^r(x,y)&=&G_{\phi}^t-G_{\phi}^{<}=G_{\phi}^{>}-
G^{\overline{t}}_{\phi}, 
\:\:\:\: G_{\phi}^a(x,y)=G^t_{\phi}-G^{>}_{\phi}=G_{\phi}^{<}-
G^{\overline{t}}_{\phi},
\end{eqnarray}
where the last two Green's functions are the retarded and advanced 
Green's functions respectively and $\theta(x,y)\equiv\theta(t_x-t_y)$ 
is the step function.  

For a generic fermion field $\psi$ the six 
different propagators are analogously defined as
\begin{eqnarray}
\label{def2}
G^{>}_{\psi}\left(x, y\right)&=&-G^{-+}_{\psi}\left(x, y\right)=-i\langle
\psi(x)\overline{\psi} (y)\rangle,\nonumber\\
G^{<}_{\psi}\left(x,y\right)&=&-G^{+-}_{\psi}\left(x, y\right)=+i\langle
\overline{\psi}(y)\psi(x)\rangle,\nonumber\\
G^{t}_{\psi} (x,y)&=& G^{++}_{\psi}\left(x, y\right)
=\theta(x,y) G^{>}_{\psi}(x,y)+
\theta(y,x) G^{<}_{\psi}(x,y),\nonumber\\
G^{\overline{t}}_{\psi} (x,y)&=& G^{--}_{\psi}\left(x, y\right)=
\theta(y,x) G^{>}_{\psi}(x,y)+
\theta(x,y) G^{<}_{\psi}(x,y),\nonumber\\
G^r_{\psi}(x,y)&=&G^{t}_{\psi}-G^{<}_{\psi}=G^{>}_{\psi}
-G^{\overline{t}}_{\psi}, \:\:\:\: G^a_{\psi}(x,y)=G^{t}_{\psi}-
G^{>}_{\psi}=G^{<}_{\psi}-G^{\overline{t}}_{\psi}.
\end{eqnarray}
From the definitions of the Green's functions, one can see
that  the hermiticity properties

\begin{equation}
\label{prop}
\left(i\gamma^0 G_\psi(x,y)\right)^\dagger=i\gamma^0  G_\psi(y,x), \,\,\,
\left(i G_\phi(x,y)\right)^\dagger=i G_\phi(y,x).
\end{equation}
are satisfied.
When computing a loop diagram, one has to assign to the  interaction points
a plus or a minus sign in all possible manners and sum all the possible
diagrams, taking into account that, by definition, vertices with  a minus sign
must be multiplied by $-1$.
 
For equilibrium phenomena, the brackets $\langle \cdots\rangle$ 
imply a thermodynamic average over all the possible states 
of the system. For homogeneous systems in equilibrium, the 
Green's functions
depend only upon the difference of their arguments $(x,y)=(x-y)$ 
and there is no dependence upon $(x+y)$;  
for systems out of equilibrium, the 
definitions (\ref{def1}) and (\ref{def2}) 
have a different meaning. The concept of thermodynamic averaging  
is now ill-defined. Instead, the brackets mean the need to 
average over all the available states of the system for the non-equilibrium 
distributions. Furthermore, the arguments of the Green's functions 
$(x,y)$ are  not usually given as the difference $(x-y)$. 
For example, non-equilibrium could be caused 
by transients which make the Green's functions
depend upon $(t_x,t_y)$ rather than $(t_x-t_y)$. 

For interacting systems,
whether in equilibrium or not, one must 
define and calculate self-energy functions. 
Again, there are six of them: $\Sigma^{t}$, 
$\Sigma^{\overline{t}}$, $\Sigma^{<}$, $\Sigma^{>}$, 
$\Sigma^r$ and $\Sigma^a$. The same 
relationships exist among them as for the 
Green's functions in  (\ref{def1}) and (\ref{def2}), such as
\begin{equation}
\Sigma^r=\Sigma^{t}-\Sigma^{<}=\Sigma^{>}-\Sigma^{\overline{t}}, 
\:\:\:\:\Sigma^a=\Sigma^{t}-\Sigma^{>}=\Sigma^{<}-\Sigma^{\overline{t}}. 
\end{equation}
The self-energies are incorporated into the Green's 
functions through the use of  Dyson's equations. 
A useful notation may be introduced which expresses 
four of the six Green's functions as the elements of 
two-by-two matrices \cite{craig}

\begin{equation}
\widetilde{G}=\left(
\begin{array}{cc}
G^{t} & \pm G^{<}\\
G^{>} & - G^{\overline{t}}
\end{array}\right), \:\:\:\:
\widetilde{\Sigma}=\left(
\begin{array}{cc}
\Sigma^{t} & \pm \Sigma^{<}\\
\Sigma^{>} & - \Sigma^{\overline{t}}
\end{array}\right),
\end{equation}
where the upper signs refer to the bosonic case and the lower signs 
to the fermionic case. For systems either in equilibrium or in non-equilibrium, 
Dyson's equation is most easily expressed by using the matrix notation
\begin{equation}
\label{d1}
\widetilde{G}(x,y)=\widetilde{G}^0(x,y)+\int d^4 z_1
\int d^4 z_2 \: \widetilde{G}^0(x,z_1)
\widetilde{\Sigma}(z_1,z_2)\widetilde{G}(z_2,y),
\end{equation}
where the superscript ``0'' on the Green's functions means 
to use those for noninteracting system.   It is useful 
to notice that Dyson's equation can be written in an 
alternative form, instead of  (\ref{d1}), with $\widetilde{G}^0$ 
on the right in the interaction terms,
\begin{equation}
\label{d2}
\widetilde{G}(x,y)=\widetilde{G}^0(x,y)+\int\: d^4z_3\:\int d^4z_4\: 
\widetilde{G}(x,z_3)
\widetilde{\Sigma}(z_3,z_4)\widetilde{G}^0(z_4,y).
\end{equation}
Eqs.~(\ref{d1}) and (\ref{d2}) are the 
starting points to derive the quantum Boltzmann equations
describing the time evolution of the 
CP-violating particle density asymmetries.

For a generic complex scalar field 
we will adopt the real-time propagator in the form 
$G^{t}_\phi(\k,t_x-t_y)$ in terms of the spectral
function $\rho_\phi(\k,k_0)$ \cite{buchmuller} (for the equilibrium analog, see \cite{ww})
\begin{eqnarray}
G^{t}_\phi(\k,t_x-t_y)&=&\int_{-\infty}^{+\infty}\:
\frac{d k^0}{2\pi}\:e^{-i k^0(t_x-t_y)}\:\rho_\phi(\k,k^0)\nonumber\\
&\times&\left\{
\left[1+f_\phi(k^0)\right]\theta(t_x-t_y)+f_\phi(k^0)\theta(t_y-t_x)\right\},
\label{rho}
\end{eqnarray}
where $f_\phi(k^0)$ represents the scalar distribution function. 
Notice that the expression (\ref{rho}) is only strictly correct for  free 
Green's functions.
To account for
interactions with the surrounding particles of the thermal bath,
particles must be
substituted by quasiparticles,  dressed propagators are to be adopted
and  self-energy corrections to the propagator modify the dispersion 
relations by  introducing a finite width $\Gamma_\phi(k)$. 
In the limit of small decay width,  the spectral function is expressed by
\begin{equation}
\rho_\phi(\k,k^0)=i\:\left[\frac{1}{(k^0+i\epsilon+ i\Gamma_\phi)^2
-\omega_\phi^2(k)}-
\frac{1}{(k^0-i\epsilon-i\Gamma_\phi)^2-\omega_\phi^2(k)}\right],
\end{equation}
where $\omega_\phi^2(k)=\k^2 +M_\phi^2(T)$ and 
$M_\phi(T)$ is the thermal mass.   
Performing the integration over $k^0$ and picking up 
the poles of the spectral function (which is valid for 
quasi-particles in equilibrium or very close to equilibrium), 
one gets 
\bea
G_\phi^{>}(\k,t_x-t_y)&=&\frac{i}{2\omega_\phi}
\left\{
\left[1+f_\phi(\omega_\phi-i\Gamma_\phi)\right]\:
e^{-i(\omega_\phi-i\Gamma_\phi)(t_x-t_y)}\right.\nonumber\\
&+&\left.
\overline{f}_\phi(\omega_\phi+i\Gamma_\phi)\:
{\rm
e}^{i(\omega_\phi+i\Gamma_\phi)(t_x-t_y)}\right\},\nonumber\\
G_\phi^{<}(\k,t_y-t_x)&=&\frac{i}{2\omega_\phi}\left\{
f_\phi(\omega_\phi+i\Gamma_\phi)\:e^{-i(\omega_\phi-i\Gamma_\phi)(t_x-t_y)}
\right.\nonumber\\
&+&\left.\left[1+\overline{f}_\phi(\omega_\phi-i\Gamma_\phi)\right]\:
e^{i(\omega_\phi+i\Gamma_\phi)(t_x-t_y)}\right\},
\label{a}
\eea
where $f_\phi$ and $\overline{f}_\phi$ denote the distribution function of
the scalar particles and antiparticles, respectively. The expressions (\ref{a}) 
are valid for $t_x-t_y>0$.

Similarly, for a generic fermion $\psi$, 
we adopt the real-time propagator in the form  
$G^{t}_\psi(\k,t_x-t_y)$ in terms of the spectral
function $\rho_\psi(\k,k_0)$ \cite{buchmuller} (for the equilibrium analog, see \cite{ww})
\bea
G^{t}_\psi(\k,t_x-t_y)&=&\int_{-\infty}^{+\infty}\:
\frac{d k^0}{2\pi}\:e^{-i k^0(t_x-t_y)}\:\rho_\psi(\k,k^0)\nonumber\\
&\times&\left\{
\left[1-f_\psi(k^0)\right]\theta(t_x-t_y)-f_\psi(k^0)\theta(t_y-t_x)\right\},
\label{rho1}
\eea
where $f_\psi(k^0)$ represents the fermion distribution function.
Again, particles must be
substituted by quasiparticles,  dressed propagators are to be adopted
and 
 self-energy
corrections to
the propagator modify the dispersion relations by 
introducing a finite width $\Gamma_\psi(k)$. 
Equilibrium fermionic dispersion
relations
are highly nontrivial \cite{weldon} and here we will adopt 
relatively simple expressions
for the fermionic spectral functions holding in the limit
in which the damping rate is smaller than the real part of the  self-energy
of the fermionic excitation \cite{henning}. For a fermion with 
chiral mass $m_\psi$, it reads
\begin{equation}\fl\qquad
\rho_\psi({\bf k},k^0) =  i\left(\not  k +m_\psi\right)
\left[\frac{1}{(k^0+i\epsilon+ i
\Gamma_{\psi})^2-\omega_{\psi}^2(k)}-
\frac{1}{(k^0-i\epsilon-i\Gamma_{\psi})^2
-\omega_{\psi}^2(k)}\right],
\label{rofermion}
\end{equation}
where  $\omega_{\psi}^2(k)={\bf k}^2 +
M_{\psi}^2(T)$ and $M_{\psi}(T)$
is the effective thermal mass of the fermion in the plasma (not a
chiral mass). 
We reiterate the fact  that at finite temperature the dispersion relation is
in fact more complicated than (\ref{rofermion}). However, the latter
suffices for our goals. 
Performing the integration over $k^0$ and picking up 
the poles of the spectral function (which is valid for 
quasi-particles in equilibrium or very close to equilibrium), 
one gets 
\begin{eqnarray}
G_\psi^{>}(\k,t_x-t_y)&=&-\frac{i}{2\omega_\psi}
\left\{
\left(\not k+ m_\psi\right) 
\left[1-f_\psi(\omega_\psi-i\Gamma_\psi)\right]\:
e^{-i(\omega_\psi-i\Gamma_\psi)(t_x-t_y)}\right.\nonumber\\
&+&\left.\gamma^0\left(\not k- 
m_\psi\right) \gamma^0
\overline{f}_\psi(\omega_\psi+i\Gamma_\psi)\:
e^{i(\omega_\psi+i\Gamma_\psi)(t_x-t_y)}\right\},\nonumber\\
G_\psi^{<}(\k,t_y-t_x)&=&\frac{i}{2\omega_\psi}\left\{
\left(\not k+ m_\psi\right)
f_\psi(\omega_\psi+i\Gamma_\psi)\:
e^{-i(\omega_\psi-i\Gamma_\psi)(t_x-t_y)}\right.\nonumber\\
&+&\left.\gamma^0\left(\not k- m_\psi\right)
\gamma^0
\left[1-\overline{f}_\psi(\omega_\psi-i\Gamma_\psi)\right]\:
e^{i(\omega_\psi+i\Gamma_\psi)(t_x-t_y)}\right\},  
\label{b}
\end{eqnarray}
where $k^0=\omega_\psi$ and  
$f_\psi, \overline{f}_\psi$ denote the distribution function of
the fermion particles and antiparticles, respectively. 
The expressions (\ref{b}) are valid for $t_x-t_y>0$.

The above definitions
hold for the Higgs and lepton doublets (after inserting the
chiral left-handed projector $P_L$), as well as for the Majorana
RH neutrinos, for which one has to assume identical  particle and antiparticle
distribution functions and insert the inverse of the charge conjugation matrix 
$C$ in the dispersion relation.


\section{The quantum Boltzmann equation for the lepton asymmetry: generalities}
\label{qbegeneral}

Since we are interested in the lepton asymmetry, 
we define the  fermionic CP-violating current of the lepton doublet 
$\ell_i$ as
\begin{equation}
\langle J_{\ell_i}^\mu(x) \rangle \equiv  \langle 
\bar{{\ell_i}}(x)\gamma^\mu {\ell_i}(x)
\rangle\equiv \left( n_{{\cal L}_i}(x), 
\vec{J}_{{\cal L}_i}(x)\right).
\end{equation} 
The zero-component of this current 
$n_{{\cal L}_i}$ represents the number density of particles minus 
the number density of antiparticles and is therefore the 
relevant quantity  for leptogenesis.

We want to find a couple of  equations of motion for 
the interacting fermionic Green's function $\widetilde{G}_{\ell_i}(x,y)$ 
when the system is not in equilibrium. Such  equations  may be found  
by applying  the operators 
$i\stackrel{\rightarrow}{\not  \partial}_x$
and $i\stackrel{\leftarrow}{\not  \partial}_y$  
on both sides of  Eqs. (\ref{d1}) and (\ref{d2}), respectively. 
We find
\begin{eqnarray}
\label{c}
i\stackrel{\rightarrow}{\not\partial}_x\widetilde{G}_{\ell_i}(x,y)&=&
\delta^{(4)}(x,y)I+
\int\:d^4 z \widetilde{\Sigma}_{\ell_i}(x,z)\widetilde{G}_{\ell_i}(z,y),
\nonumber\\
\widetilde{G}_{\ell_i}(x,y)i\stackrel{\leftarrow}{\not  \partial}_y&=& 
-\delta^{(4)}(x,y) I
-\int\:d^4 z \widetilde{G}_{\ell_i}(x,z)\widetilde{\Sigma}_{\ell_i}(z,y), 
\end{eqnarray}
where $I$ is the $4\times 4$ identity matrix. 
We can  now  take the trace over the spinorial indices of  both sides of 
the equations, sum up  the two equations above  and 
finally extract the equation 
of motion for the Green's function $G^{>}_{{\ell_i}}$
\begin{eqnarray}
\fl\qquad
\label{v}
{\rm Tr} \left\{\left[i\stackrel{\rightarrow}{\not  \partial}_x + 
i\stackrel{\leftarrow}{\not  \partial}_y\right]
G^{>}_{{\ell_i}}(x,y)\right\}&=& 
\int\:d^4 z\:{\rm Tr}\left[\Sigma^{>}_{\ell_i}(x,z)
G^t_{\ell_i}(z,y)-\Sigma^{\overline{t}}_{\ell_i}(x,z)
G^{>}_{\ell_i}(z,y)\right.\nonumber\\
&-&
\left. G^{>}_{\ell_i}(x,z)\Sigma^t_{\ell_i}(z,y)+G^{\overline{t}}_{\ell_i}(x,z)
\Sigma^{>}_{\ell_i}(zy,y)\right].
\end{eqnarray}
Since we want to compute the average of observables at equal time, 
we will identify the variables $x$ and $y$. 
Therefore, it turns out useful to  define  a center-of-mass coordinate system
\begin{equation}
\label{dd}
X\equiv(t,\vec{X})\equiv\frac{1}{2}(x+y),\:\:\:\: (\tau,\vec{r})\equiv x-y.
\end{equation}
The notation of the Green's function is altered in  these center-of-mass 
coordinates
\begin{equation}
\fl\qquad
G^{>}_{{\ell_i}}(x,y)=G^{>}_{{\ell_i}}(\tau,\vec{r},t,\vec{X})=i\langle 
{\ell_i}\left(t-\frac{1}{2}\tau, \vec{X}-\frac{1}{2}
\vec{r}\right)\overline{{\ell_i}}\left(t+\frac{1}{2}\tau, \vec{X}+\frac{1}{2}
\vec{r}\right)\rangle.
\end{equation}
Making use of the center-of-mass coordinate system,  
we can work out the left-hand side of Eq.~(\ref{v})
\begin{eqnarray}
&&\left.{\rm Tr}\left[i\stackrel{\rightarrow}{\not  \partial}_x 
G^{>}_{\ell_i}(t,\vec{X},\tau,\vec{r})+
G^{>}_{\ell_i}(t,\vec{X},\tau,\vec{r})i
\stackrel{\leftarrow}{\not  \partial}_y\right]
\right|_{\tau=\vec{r}=0}\nonumber\\
&=&\left.i\left(  \partial^x_\mu + \partial^y_\mu\right) i \langle
\overline{{\ell_i}}\gamma^\mu {\ell_i}\rangle\right|_{\tau=\vec{r}=0} 
-\frac{\partial}{\partial X^\mu} \langle
\overline{{\ell_i}}(X)\gamma^\mu 
{\ell_i}(X)\rangle=-\frac{\partial}{\partial X^\mu} J^\mu_{\ell_i}.
\end{eqnarray}

The next step is to employ the definitions in (\ref{def2}) to 
express the time-ordered functions $G^{t}_{{\ell_i}}$, 
$G^{\overline{t}}_{{\ell_i}}$, 
$\Sigma ^t_{\ell_i}$, and 
$\Sigma^{\overline{t}}_{{\ell_i}}$ in terms of $G^{<}_{{\ell_i}}$, 
$G^{>}_{{\ell_i}}$, 
 $\Sigma^{<}_{{\ell_i}}$ and  $G^{>}_{{\ell_i}}$. 
The time integrals are separated into whether
$t_z>t$ or $t_z<t$ and the right-hand side of Eq. (\ref{v}) reads,
after setting $x=y$
\begin{eqnarray}
&&\int\: d^3 z\:\int_{0}^{t} dt_z \left[\theta(t-t_z)
\left(\Sigma^{>}_{\ell_i} G^{<}_{\ell_i}+G^{<}_{\ell_i}
\Sigma^{>}_{\ell_i}-
\Sigma^{<}_{\ell_i} G^{>}_{\ell_i}-G^{>}_{\ell_i}
\Sigma^{<}_{\ell_i}
\right)\right.\nonumber\\
&&+\left. \theta(t_z-t)
\left(\Sigma^{>}_{{\ell_i}} G^{>}_{{\ell_i}}+G^{>}_{{\ell_i}}
\Sigma^{>}_{{\ell_i}}-
\Sigma^{>}_{{\ell_i}} G^{>}_{{\ell_i}}-G^{>}_{{\ell_i}}
\Sigma^{>}_{{\ell_i}}\right)\right].
\end{eqnarray}
The terms with $t_z>t$ all cancel,  leaving
\begin{eqnarray}
\label{aa}
 \frac{\partial n_{{\cal L}_i}(X)}{\partial t}
&=&-\int\: d^3 z\:\int_{0}^{t} dt_z\:{\rm Tr}
\left[\Sigma^{>}_{{\ell_i}}(X,z) G^{<}_{{\ell_i}}(z,X)
-  G^{>}_{{\ell_i}}(X,z) \Sigma^{<}_{{\ell_i}}(z,X) \right.\nonumber\\  
&+&\left.  G^{<}_{{\ell_i}}(X,z)\Sigma^{>}_{{\ell_i}}(z,X)-\Sigma^{<}_{{\ell_i}}(X,z) 
G^{>}_{{\ell_i}}(z,X)\right].
\end{eqnarray}
This is the quantum Boltzmann  equation describing the time 
evolution of a lepton  number asymmetry $n_{{\cal L}_i}$. 
The initial conditions are set at $t=0$ when the lepton 
asymmetries and the RH neutrino abundances are assumed to be vanishing.
All the information 
regarding lepton  number violating interactions 
and CP-violating sources  are  stored in the self-energy $\Sigma_{\ell_i}$. 
The kinetic  Eq.~(\ref{aa}) has an obvious interpretation in terms of 
gain and loss processes.   
What is unusual, however,   is the presence of the integral over 
the time where the theta function ensures that
the dynamics is causal. The equation is manifestly non-Markovian. Only  
the assumption that the relaxation timescale of the particle asymmetry 
is much longer than the timescale of the non-local kernels leads to 
a Markovian description. A further approximation, {\it i.e.} taking the 
upper limit of the 
time integral to $t\rightarrow \infty$,  leads to the familiar Boltzmann  
equation. 
The physical interpretation of the integral over the past history of 
the system is straightforward: it leads to the typical ``memory'' 
effects which are observed in quantum transport theory 
\cite{dan, henning, boyanovsky}. 
In  the classical kinetic theory the ``scattering term'' 
does not include any integral over the past history of the 
system which is equivalent to assume that any collision in the plasma  
does not depend upon the previous ones. On the contrary,   
quantum distributions possess strong memory effects and the thermalization 
rate obtained from the quantum transport theory may be 
substantially longer than the one obtained from 
the classical kinetic theory. The very same effects 
play a fundamental role in electroweak baryogenesis \cite{riobau}.


\section{The quantum Boltzmann equation for the right-handed neutrinos}
\label{qberhneutrino}

To get familiar with the out-of-equilibrium technique, we 
derive the quantum Boltzmann equation for the abundance of RH neutrinos.

The  Lagrangian we consider  consists of the SM one plus 
three RH neutrinos 
$N_{\alpha}$ ($\alpha=1,2,3$), with Majorana masses $M_\alpha$.  
The interactions among RH neutrinos, Higgs doublets $H$, lepton doublets  
$\ell_i$ and singlets $e_i$  ($i=e,\mu,\tau$) 
are described by the Lagrangian
\begin{equation}
{\mathscr L}_{\rm int}=\lambda_{\alpha i} N_\alpha
\ell_i H+h_{i}
\bar{e}_{i} \ell_i H^c +{1\over 2}M_\alpha N_\alpha N_\alpha + {\rm h.c.}\,,
\label{lagr_flav}
\end{equation}
with summation over repeated indeces.
The Lagrangian is written in the mass eigenstate basis of RH neutrinos and
charged leptons.

We focus here on the dynamics of the lightest RH neutrino $N_1$. 
To find its quantum Boltzmann 
equation we start from Eq.~(\ref{d1}) for the Green's function $G^<_{N_1}$ 
of the RH neutrino
$N_1$

\begin{eqnarray}
\fl\qquad
\label{vv}
\left(i\stackrel{\rightarrow}{\not  \partial}_x -M_1\right)
G^{<}_{N_1}(x,y)&=&- 
\int d^4 z \,\left[-\Sigma^{t}_{N_1}(x,z)
G^<_{N_1}(z,y)+\Sigma^{<}_{N_1}(x,z)
G^{\overline{t}}_{N_1}(z,y)
\right]\nonumber\\
&=&\int\: d^3 z\:\int_{0}^{t}\: dt_z\:
\left[\Sigma^{>}_{N_1}(x,z)
G^<_{N_1}(z,y)-\Sigma^{<}_{N_1}(x,z)
G^{>}_{N_1}(z,y)
\right].\nonumber\\
&&
\end{eqnarray}
On the left-hand side of 
this equation we perform a number of operation. We first go  
to the center-of-mass coordinates and 
perform a Fourier transform over the spatial internal coordinates
$\vec{r}$. We then insert  the expression in Eq.~(\ref{b}) for the corresponding
RH neutrino Green's function. The real part of 
 the left-hand side of Eq. (\ref{vv})
gives, after setting $x=y$, projecting onto  the positive frequencies
and taking the trace over the spinorial indeces

\begin{equation}
{\rm Re}\left[{\rm Tr}
\left(\frac{i}{2}\stackrel{\rightarrow}{\not  \partial}_X
\frac{i}{2\omega_{N_1}}\left(\not k+ M_1\right)f_{N_1}\right)\right]=-
\frac{\partial f_{N_1}({\bf k},t) }{\partial t},
\end{equation}

\begin{figure}[t]
\centering
\includegraphics{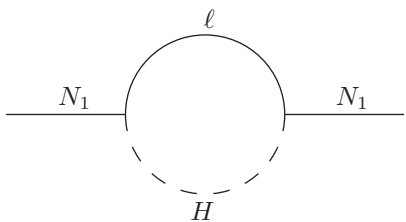}
\caption{One-loop self-energy of the lightest RH neutrino.}
\label{RHself}
\end{figure}

The self-energy of the RH neutrino is given diagrammatically in Fig.~\ref{RHself} (where
$\ell$ indicates the generic lepton doublet in the loop) and reads

\begin{equation}
\Sigma^{>,<}_{N_1}(x,y)=i\,G^{>,<}_{H}(x,y)G^{>,<}_{{\ell}}(x,y).
\end{equation}
Inserting in the right-hand side of Eq.~(\ref{vv}) the expressions in Eqs.~(\ref{a}) 
and 
(\ref{b}) for the Higgs and lepton doublet Green's functions and taking the
real part of it and the trace of the spinorial indices, we find

\begin{eqnarray}
\fl\qquad
\frac{\partial f_{N_1}({\bf k},t) }{\partial t}&=&
-2
\int_{0}^{t} dt_z\, \int\frac{d^3{\bf p}}{(2\pi)^3}\, 
\frac{1}{2\omega_{\ell}({\bf p})}\frac{1}{2\omega_H({\bf k}-{\bf p})}
\frac{1}{\omega_{N_1}({\bf k})}
\left|{\cal M}(N_1\rightarrow {\ell} H)\right|^2
\nonumber\\
&& \times \left[
f_{N_1}({\bf k},t)(1-f_{{\ell}}({\bf p},t))(1+f_{H}({\bf k}-{\bf p},t))
\right.\nonumber\\
&&\left.
-f_{{\ell}}({\bf p},t)f_{H}({\bf k}-{\bf p},t)(1-f_{N_1}({\bf k},t))
\right]\nonumber\\
 && \times \cos\left[\left(\omega_{N_1}({\bf k})-
\omega_{\ell}({\bf p})-
\omega_H({\bf k}-{\bf p})\right)(t-t_z)\right]\nonumber\\
&\simeq& -2
\int_{0}^{t} dt_z\, \int\frac{d^3{\bf p}}{(2\pi)^3}\, 
\frac{1}{2\omega_{\ell}({\bf p})}\frac{1}{2\omega_H({\bf k}-{\bf p})}
\frac{1}{\omega_{N_1}({\bf k})}\left|{\cal M}(N_1\rightarrow {\ell} H)
\right|^2\nonumber\\
&& \times\left(
f_{N_1}({\bf k},t)-f^{\rm eq}_{{\ell}}({\bf p})f^{\rm eq}_{H}({\bf k}-
{\bf p})
\right)\nonumber\\
 && \times\cos\left[\left(\omega_{N_1}({\bf k})-
\omega_{\ell}({\bf p})-
\omega_H({\bf k}-{\bf p})\right)(t-t_z)\right].
\label{cy}
\end{eqnarray}

We have made the assumption that the relaxation timescale for the
distribution functions are longer than the timescale
of the non-local kernels so that they can be extracted out of the
time integral. This allows   to think the distributions
as functions of the center-of-mass time only.
We have set to zero
the damping rates of the particles in Eq.~(\ref{b}) and
retained only those cosines giving  rise to 
energy delta functions
that can be satisfied\footnote{For simplicity, 
we neglect here the fact that thermal effects may
kinematically open new channels beyond those at zero temperature, see 
\cite{lept}.}. 
Under these assumptions, the distribution
function may be taken out of the time integral, leading -- at large times --
to the so-called
Markovian description. The kinetic equation (\ref{cy})
has an obvious interpretation in terms
of gain minus loss processes, but the retarded
time integral and the cosine function replace the familiar energy conserving
delta functions. In the second passage, we have also made the 
usual assumption  that
all  distribution functions are smaller than unity 
and that those of the Higgs and lepton doublets are in 
equilibrium and much smaller than 
unity, $f_{\ell} f_H\simeq f^{\rm eq}_{{\ell}}f^{\rm eq}_{H}$. Elastic 
scatterings are typically fast enough to keep kinetic equilibrium.
For any distribution function we may write $f=(n/n^{\rm eq})f^{\rm eq}$,
where $n$ denotes the total number density.
Therefore, Eq.~(\ref{cy}) can be re-written as

\begin{eqnarray}
\frac{\partial n_{N_1}}{\partial t}&=&
-\langle \Gamma_{N_1}(t)\rangle n_{N_1}+\langle 
\widetilde{\Gamma}_{N_1}(t)\rangle n^{\rm eq}_{N_1},\nonumber\\
\langle\Gamma_{N_1}(t)\rangle&=&\int_{0}^{t} dt_z
\int\frac{d^3{\bf k}}{(2\pi)^3}
\frac{f^{\rm eq}_{N_1}}{n^{\rm eq}_{N_1}}
\,\Gamma_{N_1}(t),\nonumber\\
\Gamma_{N_1}(t)&=&
2
\int\frac{d^3{\bf p}}{(2\pi)^3}\, 
\frac{\left|{\cal M}(N_1\rightarrow {\ell} 
H)\right|^2}{2\omega_{\ell} 2\omega_H\omega_{N_1}}
\cos\left[\left(\omega_{N_1}-
\omega_{\ell}-
\omega_H\right)(t-t_z)\right],  \nonumber\\
\langle\widetilde{\Gamma}_{N_1}(t)\rangle&=&\int_{0}^{t} dt_z
\int\frac{d^3{\bf k}}{(2\pi)^3}
\frac{f^{\rm eq}_{N_1}}{n^{\rm eq}_{N_1}}
\,\widetilde{\Gamma}_{N_1}(t),\nonumber\\
\widetilde{\Gamma}_{N_1}(t)&=&
2
\int\frac{d^3{\bf p}}{(2\pi)^3}\, 
\frac{f^{\rm eq}_{\ell}f^{\rm eq}_H}{f^{\rm eq}_{N_1}}
\frac{\left|{\cal M}(N_1\rightarrow {\ell} 
H)\right|^2}{2\omega_{\ell} 2\omega_H\omega_{N_1}}
\cos\left[\left(\omega_{N_1}-
\omega_{\ell}-
\omega_H\right)(t-t_z)\right],  \nonumber\\
\label{ddd}
\end{eqnarray}
where $\langle \Gamma_{N_1}(t)\rangle$ is the time-dependent 
thermal average of the
Lorentz-dilated decay width. 
Integrating over
large times, $t\rightarrow \infty$, thereby replacing the
cosines by energy conserving delta functions \cite{boyanovsky},

\begin{equation}
\int_{0}^{\infty}dt_z\,\cos\left[\left(\omega_{N_1}-
\omega_{\ell}-
\omega_H\right)(t-t_z)\right]=\pi\delta\left(
\omega_{N_1}-
\omega_{\ell}-
\omega_H
\right),
\end{equation}
we find that the two averaged rates $\langle\Gamma_{N_1}\rangle$ and
$\langle\widetilde{\Gamma}_{N_1}\rangle$ coincide and we recover 
the usual classical Boltzmann equation for the RH distribution
function

\begin{eqnarray}
\frac{\partial n_{N_1}}{\partial t}&=&-\langle\Gamma_{N_1}\rangle\left(
n_{N_1}-n^{\rm eq}_{N_1}
\right),\nonumber\\
\langle\Gamma_{N_1}\rangle&=&
\int\frac{d^3{\bf k}}{(2\pi)^3}
\frac{f^{\rm eq}_{N_1}}{n^{\rm eq}_{N_1}}
\int\frac{d^3{\bf p}}{(2\pi)^3}\, 
\frac{\left|{\cal M}(N_1\rightarrow {\ell} H)
\right|^2}{2\omega_{\ell}2\omega_H\omega_{N_1}}\,(2\pi)\delta\left(
\omega_{N_1}-
\omega_{\ell}-
\omega_H\right).\nonumber\\
&&
\end{eqnarray}
Taking the time interval to infinity, namely implementing Fermi's golden rule,
results in neglecting memory effects, which in turn results only
in on-shell processes contributing to the rate equation. The main difference
between the classical and the quantum Boltzmann equations can be traced to
memory effects and to the fact that the time evolution
of the distribution function is
non-Markovian. The memory of the past time evolution 
translates into off-shell processes. It would be certainly interesting
to perform a numerical study to assess the impact of the memory effects
onto the final baryon asymmetry.


\section{The quantum Boltzmann equation for the lepton asymmetry and the 
CP asymmetry}
\label{qbeleptonasym}

Our goal is now to compute the right-hand side of the  Eq.~(\ref{aa})
describing the evolution of the lepton asymmetry following  the 
CTP approach. We start with the CP
asymmetry source term.
We will see that  the CP asymmetry  manifests memory effects, 
evolves in time and at large times  may  resemble
the usual CP asymmetry expression
existing  in the literature only if certain conditions are satisfied.

\begin{figure}[t]
\centering
\includegraphics{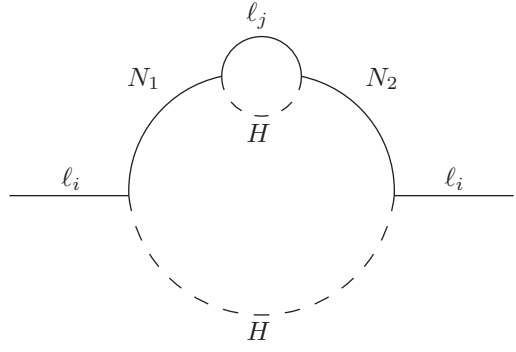}
\caption{The wave-diagram contributing to the two-loop self-energy of the
lepton doublet.}
\label{wavediagram}
\end{figure}

As we have 
already mentioned, Eq.~(\ref{aa}) contains the information about all possible 
interesting processes for leptogenesis, {\it e.g.} $\Delta L=1$ 
inverse decays, $\Delta L=2$ scatterings and so on. 
To extract the CP asymmetry, we first consider the
``wave''-diagram contribution to 
the lepton doublet ${\ell}_i$ (see Fig.~\ref{wavediagram}). 
From the previous discussion, we know that this diagram is in fact a sum of
diagrams obtained  assigning to the  interaction points
a plus or a minus sign in all possible manners, 
taking into account that, by definition, vertices with  a minus sign
must be multiplied by $-1$. The sum of all diagrams
has to give rise to theta functions which ensure that the dynamics is
causal. Since baryogenesis is a process close to equilibrium and we know
that a vanishing baryon asymmetry has to be recovered when the RH neutrinos
are in equilibrium, 
we can expand
at linear order the Green's functions of the RH neutrinos
by expanding their distribution functions around equilibrium $\delta 
f_{N_1}=f_{N_1}-f^{\rm eq}_{N_1}$:
\begin{eqnarray}
\delta G^{>}_{N_1}({\bf k},t_x-t_y)&=&\delta G^{<}_{N_1}({\bf k},t_y-t_x)
\nonumber\\
&=&
i\,\frac{e^{-\Gamma_{N_1}\left|t_x-t_y\right|}}{2\omega_{N_1}}
\left\{
\left(\not k+ M_1\right) \:{\rm
e}^{-i\omega_{N_1}(t_x-t_y)}\right.\nonumber\\
&& \left.-\gamma^0\left(\not k- 
M_1\right) \gamma^0{\rm
e}^{i\omega_{N_1}(t_x-t_y)}\right\}\times C^{-1}\,\delta f_{N_1}\,,
\end{eqnarray}
where we have neglected the dependence of the distribution function
on the damping rate. 
Summing all possible diagrams 
contributing to the loop of Fig.~\ref{wavediagram}, after a
long,
but straightforward computation, we find, 
for instance, 
\bea
&&i\left(\Sigma^{>}_{\ell_i}(X,z)G^{<}_{\ell_i}(z,X)
\Sigma^{<}_{\ell_i}(X,z)G^{>}_{\ell_i}(z,X)\right)\nonumber\\
&=&\sum_{j=1}^3
\left(\lambda_{1i}
\lambda_{1j}\lambda^\dagger_{j2}\lambda^\dagger_{i2}\right)
\int d^4 x_1\int d^4 x_2\,\delta G^{>}_{N_1}(X,x_1)
\nonumber\\
&&\times
\left[G^{<}_{\ell_j}(x_1,x_2)G^{<}_{H}(x_1,x_2)-
G^{>}_{\ell_j}(x_1,x_2)G^{>}_{H}(x_1,x_2)
\right]\nonumber\\
&&\times \left[G^{<}_{N_2}(x_2,z)-
G^{>}_{N_2}(x_2,z)\right]\nonumber\\
&&\times\left[G^{<}_{\ell_i}(z,X)G^{<}_{H}(z,X)-
G^{>}_{\ell_i}(z,X)G^{>}_{H}(z,X)
\right]\nonumber\\
&&\times\theta(z,x_2) \theta(x_2,x_1)\theta(X,z),
\eea
where we have only retained the contribution from the RH neutrino
$N_2$, since we will focus on the resonance  case
in which $M_1$ and $M_2$ are nearly degenerate. 

Making use of the  hermiticity properties in Eq.~(\ref{prop}) and 
inserting the various propagators for the Higgs fields, the lepton doublets
and the heavy Majorana neutrinos, we find that Eq.~(\ref{aa}) 
for the lepton asymmetry - as far as the CP violating source
is concerned - is given by

\begin{eqnarray}
\fl\quad
\frac{\partial n_{{\cal L}_i}}{\partial t}
&\fl\qquad\qquad=&\;
\epsilon^{{\rm W}\,i}_{N_1}(t)
\langle\Gamma_{N_1}\rangle\left(
n_{N_1}-n^{\rm eq}_{N_1}
\right),\nonumber\\
\fl\quad
\epsilon^{{\rm W}\,i}_{N_1}(t)
&\fl\qquad\qquad=&
-\frac{4}{\langle\Gamma_{N_1}\rangle}\sum_{j=1}^3{\rm Im}
\left(\lambda_{1i}
\lambda_{1j}\lambda^\dagger_{j2}
\lambda^\dagger_{i2}\right)\nonumber\\
&\fl\qquad\qquad\times&\int_0^t dt_z\int_0^{t_z} dt_2\int_0^{t_2} dt_1
e^{-\Gamma_{N_2}(t_z-t_2)}
e^{-\left(\Gamma_{\ell_j}+\Gamma_H\right)(t_2-t_1)}
\int\frac{d^3{\bf k}}{(2\pi)^3}
\frac{f^{\rm eq}_{N_1}}{n^{\rm eq}_{N_1}}\nonumber\\
&\fl\qquad\qquad\times&
\int\frac{d^3{\bf p}}{(2\pi)^3}
\frac{1-f_{\ell_j}^{\rm eq}({\bf p})+f_{H}^{\rm eq}({\bf k}-{\bf p})
}{2\omega_{\ell_j}({\bf p})2\omega_H({\bf k}-{\bf p})
\omega_{N_1}({\bf k})}
\int\frac{d^3{\bf q}}{(2\pi)^3}
\frac{1-f_{\ell_i}^{\rm eq}({\bf q})+f_{H}^{\rm eq}({\bf k}-{\bf q})
}{2\overline{\omega}_{\ell_i}({\bf q})2\overline{\omega}_H({\bf k}
-{\bf q})
\omega_{N_2}({\bf k})}\nonumber\\
&\fl\qquad\qquad\times&
 {\rm sin}\left(\omega_{N_1}(t-t_1)+\left(
\omega_{\ell_j}+\omega_H\right)(t_1-t_2)+\omega_{N_2}(t_2-t_z)
+\left(
\overline{\omega}_{\ell_i}+\overline{\omega}_H\right)(t_z-t)
\right)\nonumber\\
&\fl\qquad\qquad\times&
{\rm Tr}\left(M_1 P_L \not p  M_2 \not q\right), 
\end{eqnarray}
where, to avoid double counting, 
we have not inserted the decay rates in the propagators of the initial
and final states and, 
for simplicity, we have assumed that the damping rates of the 
lepton doublets and the Higgs field are constant in time. This should
be a good approximation as the damping rate are to be computed 
for momenta of order of the mass of the RH neutrinos. 
As expected from first principles, we find that the
CP asymmetry is a function of time and its value at a given
instant depends upon the previous history of the system. 

Performing the time integrals and retaining only those pieces which
eventually give rise to energy-conserving delta functions in the 
Markovian limit (as in the previous Section, we do not include here
the new channels that thermal effects may eventually open), we obtain
\begin{eqnarray}
\label{eeee}
\fl\qquad
\epsilon^{{\rm W}\,i}_{N_1}(t)&=&
-\frac{4}{\langle\Gamma_{N_1}\rangle}\sum_{j=1}^3{\rm Im}
\left(\lambda_{1i}
\lambda_{1j}\lambda^\dagger_{j2}
\lambda^\dagger_{i2}\right)\nonumber\\
&\times&
\int_0^t dt_z
\frac{\cos\left[\left(\omega_{N_1}-
\overline{\omega}_{\ell_i}-
\overline{\omega}_H\right)(t-t_z)\right]}{
\left(\Gamma_{N_2}^2+(\omega_{N_2}-\omega_{N_1})^2\right)
\left((\Gamma_{\ell_j}+\Gamma_{H})^2+(\omega_{N_1}-
\omega_{\ell_j}-\omega_H
)^2\right)}
\nonumber\\
&\times&
\int\frac{d^3{\bf k}}{(2\pi)^3}
\frac{f^{\rm eq}_{N_1}}{n^{\rm eq}_{N_1}}
(\Gamma_{\ell_j}+\Gamma_{H})\Bigg(2\,(\omega_{N_2}-\omega_{N_1})
\,{\rm sin}^2 \left[\frac{(\omega_{N_2}-\omega_{N_1})t_z}{2}\right]
\nonumber\\
&-&
\Gamma_{N_2}\,{\rm sin} \left[(\omega_{N_2}-\omega_{N_1})t_z\right]
\Bigg)
\int\frac{d^3{\bf p}}{(2\pi)^3}
\frac{1-f_{\ell_j}^{\rm eq}({\bf p})+f_{H}^{\rm eq}({\bf k}-{\bf p})
}{2\omega_{\ell_j}({\bf p})2\omega_H({\bf k}-{\bf p})
\omega_{N_1}({\bf k})}
\nonumber\\
&\times&
\int\frac{d^3{\bf q}}{(2\pi)^3}
\frac{1-f_{\ell_i}^{\rm eq}({\bf q})+f_{H}^{\rm eq}({\bf k}-{\bf q})
}{2\overline{\omega}_{\ell_i}({\bf q})2\overline{\omega}_H({\bf k}
-{\bf q})
\omega_{N_2}({\bf k})}\;
{\rm Tr}\left(M_1 P_L \not p M_2 \not q\right).
\end{eqnarray}
From this expression it is already 
manifest that the typical timescale for the
building up of the coherent CP asymmetry depends crucially on
the difference in energy of the two RH neutrinos.

If we now let the upper limit of the time integral to take large values, we neglect
the memory effects, the CP asymmetry picks contribution only
from the 
on-shell processes. Taking the damping rates of the lepton doublets
equal for all the flavours and the RH neutrinos nearly at rest
with respect to the thermal bath, the CP asymmetry from the 
``wave''-diagram 
reads (now summing over all flavour indices)
\begin{eqnarray}
\epsilon^{{\rm W}}_{N_1}(t)&\simeq &-\frac{{\rm Im}
\left(\lambda\lambda^\dagger\right)^2_{12}}{\left(\lambda\lambda^\dagger
\right)_{11}\left(\lambda\lambda^\dagger
\right)_{22}}
\frac{M_1}{M_2}\Gamma_{N_2}\frac{1}{(\Delta M)^2+ 
\Gamma_{N_2}^2}\nonumber\\
&\times& \left(2\,\Delta M \,{\rm sin}^2 \left[\frac{\Delta M t}{2}\right]
-\Gamma_{N_2}\,{\rm sin} \left[\Delta M t\right]\right),
\label{ll}
\end{eqnarray}
where $\Delta M= (M_2-M_1)$. 
The CP asymmetry  (\ref{ll}) is resonantly 
enhanced when  $\Delta M\simeq \Gamma_{N_2}$ and at the resonance
point it is given by

\begin{equation}
\label{kkk}
\epsilon^{{\rm W}}_{N_1}(t)\simeq -\frac{1}{2}\frac{{\rm Im}
\left(\lambda\lambda^\dagger\right)^2 _{12}}{\left(\lambda\lambda^\dagger
\right)_{11}\left(\lambda\lambda^\dagger
\right)_{22}}
\left(1-{\rm sin} \left[\Delta M t\right]-
{\rm cos} \left[\Delta M t\right]\right),
\end{equation}
The timescale for the 
building up of the CP asymmetry is $\sim 1/\Delta M$. 
The CP asymmetry  grows starting from a vanishing value and, for
$t\gg (\Delta M)^{-1}$, it averages to the constant 
value quoted in the literature \cite{resonant}. 
This is true if the 
timescale for the other processes relevant
for leptogenesis is larger  than  $\sim 1/\Delta M$. 
In other words, one may define an ``average'' CP asymmetry

\begin{equation}
\langle \epsilon^{{\rm W}}_{N_1}\rangle=\frac{1}{\tau_{\rm p}}
\int_{t-\tau_{\rm p}}^{t}
dt^\prime\, \epsilon^{{\rm W}}_{N_1}(t^\prime),
\end{equation}
where $\tau_{\rm p}$ represents the typical timescale 
of the other processes relevant for leptogenesis, {\it e.g.}
the $\Delta L=1$ scatterings. If $\tau_{\rm p}\gg 1/\Delta M\sim 
\Gamma^{-1}_{N_2}$, 
the oscillating functions
in (\ref{kkk}) are averaged to zero and the average CP asymmetry
is given by the value used in the literature.  
However, the expression (\ref{ll}) should be
used when   $\tau_{\rm p}\lsim 1/\Delta M\sim \Gamma^{-1}_{N_2}$. 

The fact that the CP asymmetry is a function of time 
is particularly relevant  in the case in which the 
asymmetry is generated by the decays of two heavy states which are
nearly degenerate in mass and oscillate into one another with a 
timescale given by the inverse of the mass difference. 
This is the case of resonant leptogenesis \cite{resonant} 
and soft leptogenesis 
\cite{soft}.  
From Eq. (\ref{ll})
it is manifest that the CP asymmetry itself oscillates with the very same
timescale and such a dependence may or may not be
neglected depending upon the rates of the other processes in the plasma. 
If $\Gamma_{N_1}\gsim \Gamma_{N_2}$, the time dependence of the CP asymmetry
may not be neglected.

\begin{figure}[t]
\centering
\includegraphics{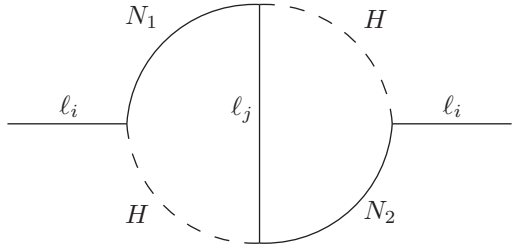}
\caption{The vertex-diagram contributing to the two-loop self-energy of the
lepton doublet.}
\label{vertexdiagriam}
\end{figure}

\begin{figure}[t]
\centering
\includegraphics[width=.48\textwidth]{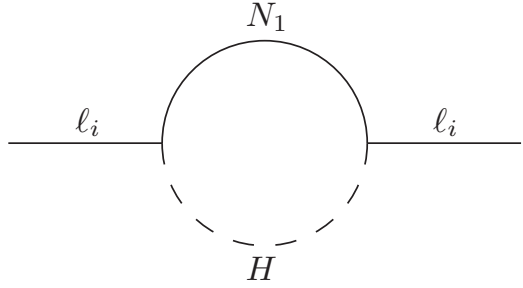}
\caption{One-loop diagram contributing to the self-energy of the
lepton doublet.}
\label{lep1loop}
\end{figure}

The expression (\ref{ll}) can also be used, once
it is divided by a factor 2 (because in the wave diagram also the charged
states of Higgs and lepton doublets may propagate) and the limit $M_2\gg M_1$
is taken, for the CP asymmetry contribution from the vertex diagram (see
Fig.~\ref{vertexdiagriam})

\begin{equation}
\epsilon^{{\rm V}}_{N_1}(t)\simeq -\frac{{\rm Im}
\left(\lambda\lambda^\dagger\right)^2_{12}}{16 \pi\left(\lambda\lambda^\dagger
\right)_{11}}
\frac{M_1}{M_2}
\left(2\, {\rm sin}^2 \left[\frac{M_2 t}{2}\right]
-\frac{\Gamma_{N_2}}{M_2}\,{\rm sin} \left[M_2 t\right]\right),
\label{lv}
\end{equation}
The timescale for this CP asymmetry is $\sim M_2$ and much larger
than any other timescale in the dynamics. Therefore, one can safely average 
over many oscillations, getting the expression present in the 
literature

\begin{equation}
\langle\epsilon^{{\rm V}}_{N_1}\rangle \simeq -\frac{{\rm Im}
\left(\lambda\lambda^\dagger\right)^2_{12}}{16 \pi\left(\lambda\lambda^\dagger
\right)_{11}}
\frac{M_1}{M_2}.
\label{lvv}
\end{equation}

Finally, 
the $\Delta L=1$ inverse decays can be computed along the same lines described
previously by considering the one-loop contribution to
the lepton doublet self-energy $\Sigma_{\ell_i}$ (see Fig.~\ref{lep1loop}). 
The equation for the
lepton asymmetry then becomes

\begin{eqnarray}
\frac{\partial n_{{\cal L}_i}}{\partial t}&=&\epsilon^{i}_{N_1}(t)
\langle\Gamma_{N_1}\rangle\left(
n_{N_1}-n^{\rm eq}_{N_1}
\right)-\langle\Gamma^{\rm ID}_{N_1}(t)\rangle\,
\frac{n^{\rm eq}_{N_1}}{2\,n^{\rm eq}_{\ell_i}}\, n_{{\cal L}_i}
\label{dddd}
\end{eqnarray}
where $\epsilon^{i}_{N_1}(t)=\epsilon^{{\rm V}\,i}_{N_1}+\epsilon^{{\rm W}\,i}_{N_1}$ 
is the total time-dependent CP asymmetry for the flavour $i$, and 
\begin{eqnarray}
\langle\Gamma^{\rm ID}_{N_1}(t)\rangle &=&
2\int_0^t dt_z \int\frac{d^3{\bf k}}{(2\pi)^3}
\frac{f^{\rm eq}_{\ell_i}f^{\rm eq}_{H}}{n^{\rm eq}_{N_1}}
\int\frac{d^3{\bf p}}{(2\pi)^3}\, 
\frac{\left|{\cal M}({\ell_i} 
H\rightarrow N_1)\right|^2}{2\omega_{\ell_i} 2\omega_H \omega_{N_1}}\nonumber\\
&\times&
 \cos\left[\left(\omega_{N_1}-
\omega_{\ell_i}-
\omega_H\right)(t-t_z)\right],
\end{eqnarray}
is the time-dependent thermal average of the inverse-decay interaction rate.


\section{Conclusions}
\label{concl}

The quantum Boltzmann equations derived in this paper can be used to
perform a thorough investigation of the impact of flavour effects
onto leptogenesis and, on more general grounds, to provide a quantitative
relation between the light neutrino properties and the final baryon asymmetry.
It would be interesting to see how large are the corrections to the
baryon asymmetry once the non-Markovian description is adopted, including
memory effects and off-shell corrections. They may lead to 
the  slowdown of the relaxation processes thus keeping
 the system out of equilibrium for longer times and therefore 
to an enhancement of the final baryon asymmetry. It would also be of interest
to see the impact of our results on the transition between one 
flavour and two flavours
as we discussed in the Introduction. 

One of the main results of our investigation is that the CP asymmetry
turns out to be a function of time and its value at a given instant
of time depends on the past history of the system, see 
for instance Eq.~(\ref{eeee}). This
result is relevant when the timescale of the evolution of the CP asymmetry
is larger than the timescale of the other processes. We have
pointed out that this is relevant  
when the asymmetry is generated by the decays of two nearly mass-degenerate
heavy states and the  resonant effects are exploited.

\ack
We thank S. Davidson for useful discussions, comments and for carefully
reading the manuscript.  
A.D.S. is supported in part by  INFN 
`Bruno Rossi' Fellowship and in part by 
the U.S. Department of Energy (D.O.E.) 
under cooperative research agreement DE-FG02-05ER41360.



\section*{References}
\begin{thebibliography}{99}
\bibitem{fy} M.~Fukugita and T.~Yanagida,
  Phys.\ Lett.\  B {\bf 174}, 45 (1986).

\bibitem{lept} G.~F.~Giudice, A.~Notari, M.~Raidal, A.~Riotto and A.~Strumia,
Nucl.\ Phys.\  B {\bf 685}, 89 (2004) [arXiv:hep-ph/0310123].


\bibitem{ogen} W.~Buchmuller, P.~Di Bari and M.~Plumacher,
Annals Phys.\  {\bf 315} (2005) 305 [arXiv:hep-ph/0401240].

\bibitem{work} A partial list:~W.~Buchmuller, P.~Di Bari and M.~Plumacher,
Nucl.\ Phys.\ B {\bf 643} (2002) 367 [arXiv:hep-ph/0205349];
J.~R.~Ellis, M.~Raidal and T.~Yanagida,
Phys.\ Lett.\ B {\bf 546} (2002) 228 [arXiv:hep-ph/0206300];
G.~C.~Branco, R.~Gonzalez Felipe, F.~R.~Joaquim and M.~N.~Rebelo,
Nucl.\ Phys.\  B {\bf 640} (2002) 202 [arXiv:hep-ph/0202030];
G.~C.~Branco, R.~Gonzalez Felipe, F.~R.~Joaquim, I.~Masina,
M.~N.~Rebelo and C.~A.~Savoy,
Phys.\ Rev.\ D {\bf 67}, 073025 (2003) [arXiv:hep-ph/0211001];
R.~N.~Mohapatra, S.~Nasri and H.~B.~Yu,
Phys.\ Lett.\  B {\bf 615} (2005) 231 [arXiv:hep-ph/0502026];
A.~Broncano, M.~B.~Gavela and E.~Jenkins,
Nucl.\ Phys.\  B {\bf 672} (2003) 163 [arXiv:hep-ph/0307058];
A.~Pilaftsis,
Phys.\ Rev.\ D {\bf 56} (1997) 5431 [arXiv:hep-ph/9707235];
E.~Nezri and J.~Orloff,
JHEP {\bf 0304} (2003) 020 [arXiv:hep-ph/0004227];
S.~Davidson and A.~Ibarra,
Nucl.\ Phys.\ B {\bf 648}, 345 (2003) [arXiv:hep-ph/0206304];
S.~Davidson,
JHEP {\bf 0303} (2003) 037 [arXiv:hep-ph/0302075];
S.~T.~Petcov, W.~Rodejohann, T.~Shindou and Y.~Takanishi,
Nucl.\ Phys.\ B {\bf 739} (2006) 208 [arXiv:hep-ph/0510404].



\bibitem{sakharov} A.D. Sakharov. \newblock  JETP Lett. {\bf 5} (1967) 24.
For a review, see A.~Riotto and M.~Trodden,
Ann.\ Rev.\ Nucl.\ Part.\ Sci.\  {\bf 49}, 35 (1999)
[arXiv:hep-ph/9901362].


\bibitem{seesaw}
P.~Minkowski,
Phys.\ Lett.\ B {\bf 67} 421 (1977); M.~Gell-Mann, P.~Ramond and R.~Slansky,
 in {\it  Supergravity}, eds.\ P.~Van Nieuwenhuizen and D.~Freedman
  (North-Holland, Amsterdam, 1979), p.~315; T.~Yanagida, in
{\it Proceedings of the Workshop on the Unified Theory and the
Baryon Number in the Universe}, eds.\ O.~Sawada and A.~Sugamoto
(KEK, Tsukuba, 1979), p.~95; S.L.~Glashow, in {\it Quarks and
Leptons}, eds.\ M.~L\'evy et al., (Plenum, 1980, New-York), p. 707;
R.N.~Mohapatra and G.~Senjanovi\'{c}, Phys.\ Rev.\ Lett.\ {\bf 44},
912 (1980).


\bibitem{Barbieri99}
R.~Barbieri, P.~Creminelli, A.~Strumia and N.~Tetradis,
  Nucl.\ Phys.\  B {\bf 575} (2000) 61
  [arXiv:hep-ph/9911315].

\bibitem{endoh}
 T.~Endoh, T.~Morozumi and Z.~h.~Xiong,
  Prog.\ Theor.\ Phys.\  {\bf 111} (2004) 123
  [arXiv:hep-ph/0308276].
 
\bibitem{davidsonetal}
A.~Abada, S.~Davidson, F.~X.~Josse-Michaux, M.~Losada and A.~Riotto,
JCAP {\bf 0604}, 004 (2006) [arXiv:hep-ph/0601083].

\bibitem{nardietal}
E.~Nardi, Y.~Nir, E.~Roulet and J.~Racker,
JHEP {\bf 0601}, 164 (2006) [arXiv:hep-ph/0601084].


\bibitem{dibari}
S.~Blanchet and P.~Di Bari,
arXiv:hep-ph/0607330.


\bibitem{davidsonetal2}
A.~Abada, S.~Davidson, A.~Ibarra, F.~X.~Josse-Michaux, M.~Losada and
A.~Riotto,
JHEP {\bf 0609}, 010 (2006) [arXiv:hep-ph/0605281].


\bibitem{antusch}
S.~Antusch, S.~F.~King and A.~Riotto,
JCAP {\bf 0611}, 011 (2006) [arXiv:hep-ph/0609038].


\bibitem{silvia1} S.~Pascoli, S.~T.~Petcov and A.~Riotto,
arXiv:hep-ph/0609125.

\bibitem{Branco:2006ce}
  G.~C.~Branco, R.~Gonzalez Felipe and F.~R.~Joaquim,
  Phys.\ Lett.\  B {\bf 645} (2007) 432
  [arXiv:hep-ph/0609297].

\bibitem{aat} S.~Antusch and A.~M.~Teixeira,
arXiv:hep-ph/0611232.

\bibitem{silvia2} S.~Pascoli, S.~T.~Petcov and A.~Riotto,
arXiv:hep-ph/0611338.

\bibitem{adsar} 
A.~De Simone and A.~Riotto,
  JCAP {\bf 0702} (2007) 005
  [arXiv:hep-ph/0611357];
S.~Blanchet, P.~Di Bari and G.~G.~Raffelt,
arXiv:hep-ph/0611337.



\bibitem{vives}
O. Vives, Phys. Rev. D {\bf 73}, 073006 (2006)
[arXiv:hep-ph/0512160].

\bibitem{Engelhard:2006yg}
G.~Engelhard, Y.~Grossman, E.~Nardi and Y.~Nir,
arXiv:hep-ph/0612187.

\bibitem{buchmuller} W.~Buchmuller and S.~Fredenhagen,
  Phys.\ Lett.\  B {\bf 483}, 217 (2000)
  [arXiv:hep-ph/0004145]. 


\bibitem{dan} P.~Danielewicz,
  Annals Phys.\  {\bf 152}, 239 (1984);
  P.~Danielewicz,
  Annals Phys.\  {\bf 152}, 305 (1984).

\bibitem{sk} J. Schwinger, J. Math. Phys. {\bf 2} 407, (1961); 
L.V. Keldysh, JETP {\bf 20} 1018, (1965); P.~M.~Bakshi and K.~T.~Mahanthappa,
  J.\ Math.\ Phys.\  {\bf 4}, 1 (1963); P.~M.~Bakshi and K.~T.~Mahanthappa,
  J.\ Math.\ Phys.\  {\bf 4}, 12 (1963); 
  K.~T.~Mahanthappa,
  Phys.\ Rev.\  {\bf 126} (1962) 329.

\bibitem{chou} 
K. Chou, Z. Su, B. Hao and L. Yu,  Phys. Rep.  {\bf 118}
1, (1985)  and references therein.

\bibitem{craig} R.A. Craig, J. Math. Phys. {\bf 9} 605,  (1968).

\bibitem{ww} N.P. Landsmann and Ch.G. van Weert, Phys. Rep. {\bf 145}
141, (1987).

\bibitem{weldon} H.A. Weldon, Phys. Rev.  {\bf D26} (1982), 
1394; V.V. Klimov, Sov. Phys. JETP  {\bf 55},  (1982) 199.
%
\bibitem{henning} P.A. Henning, Phys. Rep. {\bf 253},  235 (1995); 

\bibitem{boyanovsky}
 D.~Boyanovsky, H.~J.~de Vega, R.~Holman, S.~Prem Kumar and R.~D.~Pisarski,
  Phys.\ Rev.\  D {\bf 58}, 125009 (1998)
  [arXiv:hep-ph/9802370]



\bibitem{riobau} 
  A.~Riotto,
  Phys.\ Rev.\  D {\bf 53}, 5834 (1996) [arXiv:hep-ph/9510271]; 
  A.~Riotto,
  Nucl.\ Phys.\  B {\bf 518}, 339 (1998) [arXiv:hep-ph/9712221]; 
  A.~Riotto,
  Int.\ J.\ Mod.\ Phys.\  D {\bf 7}, 815 (1998)
  [arXiv:hep-ph/9709286];
  A.~Riotto,
  Phys.\ Rev.\  D {\bf 58}, 095009 (1998)
  [arXiv:hep-ph/9803357].

\bibitem{resonant} 
M.~Flanz, E.~A.~Paschos and U.~Sarkar,
Phys.\ Lett.\ B {\bf 345} (1995) 248 [Erratum-ibid.\ B {\bf 382}
(1996) 447] [arXiv:hep-ph/9411366];
A.~Pilaftsis,
Phys.\ Rev.\ D {\bf 56} (1997) 5431 [arXiv:hep-ph/9707235].

\bibitem{soft} G.~D'Ambrosio, G.~F.~Giudice and M.~Raidal,
  Phys.\ Lett.\  B {\bf 575}, 75 (2003)
  [arXiv:hep-ph/0308031];
  Y.~Grossman, T.~Kashti, Y.~Nir and E.~Roulet,
  Phys.\ Rev.\ Lett.\  {\bf 91} (2003) 251801
  [arXiv:hep-ph/0307081].
\end{thebibliography}
\end{document}